# Superposition of Weyl solutions: The equilibrium forces

*P.S. Letelier*[1] *and S.R. Oliveira*[2]

Departamento de Matemática Aplicada-IMECC
Universidade Estadual de Campinas
13081-970 Campinas, S.P., Brazil

Solutions to the Einstein equation that represent the superposition of static isolated bodies with axially symmetry are presented. The equations nonlinearity yields singular structures ( strut and membranes ) to equilibrate the bodies. The force on the strut like singularities is computed for a variety of situations. The superposition of a ring and a particle is studied in some detail.

[1]*e-mail: letelier@ime.unicamp.br*
[2]*e-mail: samuel@ime.unicamp.br*

# 1 Introduction

Among the most simple albeit physically meaningful solutions of the vacuum Einstein equations are the Weyl solutions [1]-[12]. They are static axially symmetric spacetimes with metric,

$$ds^2 = e^{2\psi}dt^2 - e^{-2\psi}[e^{2\nu}(dr^2 + dz^2) + r^2 d\varphi^2], \tag{1}$$

where the functions $\psi$ and $\nu$ depend on coordinates $r$ and $z$ only, and satisfy the differential equations:

$$\nabla^2 \psi \equiv \psi_{rr} + \psi_r/r + \psi_{zz} = 0, \tag{2}$$

$$d\nu[\psi] = r[(\psi_r^2 - \psi_z^2)dr + 2\psi_r \psi_z dz]. \tag{3}$$

The linearity of (2) and the particular form of (3) allows us to find explicitly solutions that represent superposition of two or more axially symmetric bodies of equal or different shapes [3][11][12]. Such a configuration is not gravitationally stable [13]. Indeed, the nonlinear terms of (3) are responsible for the apparition of gravitationally inert singular structures, "struts" and "membranes", that keep the two bodies apart making a stable configuration [14]-[20].

The simplicity of the Weyl solutions is rather deceiving since the potential $\psi$ that obeys the usual Newtonian Laplace equation has a different meaning in general relativity. The mono-polar solution of the Einstein equation, the Schwarzschild solution, is represented in these coordinates by a bar of length $2m$ [1], i.e., a "potential" whose multi-polar expansion contains all the multi-polar moments beyond the dipole [7]-[10]. Newtonian images are useful in the description of the spacetime solutions but need to be used carefully [21]-[26].

We consider superpositions of axially symmetric bodies, each body has its own singularity. The non-linear feature of Einstein equations yields other singularities in order to preserve the original static character of the spacetime. For each superposition considered we obtain also the "attraction force" between the bodies. It is the compression force on the strut-like singularity. In all the cases the Newtonian force is recovered in the appropriate limits. Recently this force has been used to make a simple analysis of the energy radiated in the head on collision of two black holes [27].

We show that in the superposition of a particular ring and a point-particle a membrane appears in the interior of the ring besides the strut between the center of the ring and the point-particle. In a previous work [20] we considered the superposition of a Chazy-Curzon particle [11][12] and an Appell "ring" [28]. It has been pointed out however that we misinterpreted Appell potential as being produced sole by a ring [29]. Actually, the Appell potential is a Newtonian potential associated to a disk of negative mass density whose rim has positive mass density. The misinterpretation of the Appell potential as representing sole a ring is frequent in the literature [3]. (See references [30][31]). In this paper we use a Newtonian potential associated to a true ring.

The paper is organized as follows. In section 2 we review some important solutions associated to isolated bodies. They are obtained from the Newtonian potential of a point particle, a rod, disks etc. In section 3 we show the general way of obtaining the compression force on the strut that appears between the superposed bodies. In section 4 we present the superposition cases: two point particles, two rods, two disks and a point particle and a particular ring. In all cases we obtain the explicit form for the compression force on the strut singularity. We close the section with some remarks on the singularities of the solutions.

## 2   Solutions associated to isolated bodies

Using simple closed form of analytical solutions one can generate others by superposition of solutions, complexification, and by parametric differentiation [21][32]-[34]. Simple asymptotically flat solutions associated to isolated bodies are:

---
[3]We could have seen our oversight in an earlier investigation – see Eqs. (2.30) of [21].

## 2.1 Chazy-Curzon particle

The solution for a point particle of mass $m$ at $z = z_0$ is given by metric (1) with functions,

$$\psi = -\frac{m}{R}, \tag{4}$$

$$\nu = -\frac{m^2 \, r^2}{2R^4}, \tag{5}$$

$$R = \sqrt{r^2 + (z - z_0)^2}. \tag{6}$$

Note that even though the metric function $\psi$ is spherically symmetric around the particle, the other metric function, $\nu$, is not. The space-time described by (4-5), known as the Chazy-Curzon solution [11][12], has a hoop-like structure [35]. The solution has a directional singularity at origin: as one approaches the origin by the $z$-axis there is no singularity [36].

## 2.2 Weyl rod

Associated to a rod centered on the origin of length $L$ and total mass $m$, we have [1] [3]

$$\psi = -\frac{m}{L} \log \left| \frac{R_1 + R_2 - L}{R_1 + R_2 + L} \right|, \tag{7}$$

$$\nu = 2 \left( \frac{m}{L} \right)^2 \log \left| \frac{(R_1 + R_2)^2 - L^2}{4 \, R_1 \, R_2} \right|, \tag{8}$$

$$R_1 = \sqrt{r^2 + \left( z - \frac{L}{2} \right)^2}, \tag{9}$$

$$R_2 = \sqrt{r^2 + \left( z + \frac{L}{2} \right)^2}. \tag{10}$$

This solution may be interpreted as the space-time of a static distorted black hole [39]-[41]. This solution is also known as $\gamma$-Weyl solution and Zipoy-Voorhees solution [23]-[25]. When $m = L$ it reduces exactly to the Schwarzschild black hole solution [7].

## 2.3 Appell disk

The solution for the Appell potential [28] can be obtained as the real part of the complexified Chazy-Curzon solution obtained by doing $z_0 \to i\,a$ [21]. We get

$$\psi = -\mathrm{Re}\left[\frac{m}{\mathsf{R}}\right], \tag{11}$$

$$\nu = -\frac{m^2\,r^2}{2}\mathrm{Re}\left[\mathsf{R}^{-4}\right] - \frac{m^2}{8a^2}\left[\frac{r^2 + z^2 + a^2}{|\mathsf{R}|^2} - 1\right], \tag{12}$$

$$\mathsf{R} = \sqrt{r^2 + (z - i\,a)^2}. \tag{13}$$

The Appell potential is associated to a disk with an inner negative mass density and a positive mass density rim; the total mass is positive. [29].

## 2.4 Morgan disk

The solution for the first member of the family of Morgan and Morgan potentials [42] can be obtained as the imaginary part of a complexified bar with $L/2 = i\,a$ [21]. We find

$$\psi = -\frac{m}{2\,i\,a}\,\log\left|\frac{\mathrm{Re}\,[\mathsf{R}] - i\,a}{\mathrm{Re}\,[\mathsf{R}] + i\,a}\right|, \tag{14}$$

$$\nu = -\frac{1}{2}\left(\frac{m}{a}\right)^2 \log\left|\frac{(\text{Re}\,[\mathsf{R}])^2 + a^2}{|\mathsf{R}|^2}\right|, \tag{15}$$

$$\mathsf{R} = \sqrt{r^2 + (z - i\,a)^2}. \tag{16}$$

This first member of the class of Morgan and Morgan disks can be used to generate the whole family [21].

## 2.5 A true ring

We obtain the solution associated to a ring by the superposition of Appell and Morgan potentials. It has been shown [29] that the membrane inside the disk in the Appell potential has exactly the same density of a Morgan disk [42] but with opposite sign. A true ring will be obtained by cancelling the inner density:

$$\psi_{\text{ring}} = \psi_{\text{Appell}} + \psi_{\text{Morgan}}, \tag{17}$$

where

$$\psi_{\text{Appell}} = -\text{Re}\left\{\frac{M/2}{R_a}\right\}, \tag{18}$$

$$\psi_{\text{Morgan}} = -\text{Im}\left\{\frac{M/2}{a}\log(\mu_a)\right\}, \tag{19}$$

and

$$R_a = \sqrt{r^2 + (z - ia)^2} \;\; ; \;\; \mu_a = ia - z + R_a. \tag{20}$$

There are several distinct solutions for Laplace equation depending on the Riemann sheets chosen for the multi valued complex functions involved in the solutions above [29]. We fix our solutions in the following way,

$$\psi_{\text{Appell}} = -\frac{M}{2\sqrt{2}} \sqrt{\frac{\sqrt{(r^2 + z^2 - a^2)^2 + 4\,z^2 a^2} + r^2 + z^2 - a^2}{(r^2 + z^2 - a^2)^2 + 4\,z^2 a^2}}, \qquad (21)$$

$$\psi_{\text{Morgan}} = -\frac{M}{2a} \arctan\left[\sqrt{\frac{2a^2}{\sqrt{(r^2 + z^2 - a^2)^2 + 4\,z^2 a^2} + r^2 + z^2 - a^2}}\right]. \qquad (22)$$

We shall always take the positive square roots. The surface mass density of the Morgan potential is given by $\sigma_{\text{Morgan}} = M/\left(4\pi a\sqrt{a^2 - r^2}\right)$. The Appell potential is associated to a disk with $-\sigma_{\text{Morgan}}$. Therefore, the superposition (17) cancels the mass density of the disk remaining exactly a ring of radius $a$ with positive total mass $M$.

In Fig. 1 we show a) The Appell potential, b) The Morgan potential, and c) The superposition potential (17) which represents a ring located on the plane $z = 0$ with $2M = a = 1$. We see that Fig. 1c is the potential of a true ring with no matter outside the line that represents the ring (a point in our figure).

Physically, the gravitational stability of this solution comes from internal tensions along the ring. We do not investigate its matter-tension contents since we are mainly interested on the external gravitational fields. See for example Refs. [37] and [38] for fluid-like and cosmic string-like rings.

The other metric function $\nu[\psi]$ can be computed using the identities of

[21]. We find

$$\nu[\psi] = -\frac{M^2}{4a^2}\left(1 + \log\left[\frac{r^2 + |\mu_a|^2}{|r^2 + \mu_a^2|}\right]\right) - \frac{M^2}{32a^2}\left(\frac{r^2 + |z - ia|^2}{|R_a|^2} - 1\right)$$
$$-\frac{M^2}{16}\text{Re}\left[\frac{r^2}{R_a^4}\right] + \frac{M^2}{2a}\text{Im}\left[\frac{r^2}{R_a(r^2 + \mu_a^2)} - \frac{\mu_a}{R_a(\mu_a - \mu_a^*)}\right]. \quad (23)$$

The first term, the integration constant $-M^2/4a^2$, was added to restore both the Minkowskian asymptotic behavior and the regularity of the metric on the $z$-axis, that is, $\nu(r = 0, z) = 0$ . The remarkable fact that this solution does not produce strut-like singularities, despite being the superposition of Appell and Morgan disks, is due to the fact that the matter contents are superimposed on the same plane. Otherwise, a strut joining the centers of the disks will be present.

# 3  Elementary flatness condition and equilibrium forces

The static asymptotically flat axisymmetric vacuum fields that we shall consider have two commuting Killing vectors. The axial symmetry is described by a space-like Killing vector field $\eta = \partial_\varphi$. Let $X = \eta^a \eta_a$ be its norm. Then the coordinate $\varphi$ has the standard periodicity $2\pi$ if the regularity condition

$$\lim_{axis} \frac{X_{,a}X^{,a}}{4X} \longmapsto 1 \quad (24)$$

is satisfied [9] . Otherwise the elementary flatness condition is violated in the vicinity of the symmetry axis and singularities are present. Note that $\eta$ vanishes at the rotation axis. For the metric (1) the condition (24) is equivalent to $\lim_{r \mapsto 0} \nu = 0$.

When $\nu(r = 0, z) \neq 0, z \in Z$ there is a strut singularity on $Z$ a subset of the $z$-axis. The compression "force" $F_z$ on a plane perpendicular to the $z$-axis can be calculated [7][43] by integrating $T_z^z$ on a disk ($t$ and $z$ fixed) centered at $r = 0$,

$$F_z = \int\int_{\text{disk}} T_z^z \, d\sigma, \tag{25}$$

where $d\sigma$ is the appropriate surface element.

We can compute a surface integral of the energy-momentum tensor through the application of the Gauss-Bonet theorem to the two-dimensional hyper-surface obtained by fixing $t$ and $z$ [44]. Its line element is conformal to

$$dr^2 + \exp(-2\nu) r^2 d\varphi^2. \tag{26}$$

A surface integral of the Gaussian curvature $K$ of this 2-d hyper-surface is given by

$$\int K d\sigma = 2\pi \left(1 - \exp\left(-\nu\left(0, z\right)\right)\right) \tag{27}$$

The Gaussian curvature is a distribution-value function. The disk is not globally flat unless $\nu(0, z) = 0$ [44][19]. The induced metric of this two-dimensional manifold is diagonal. Thus the only non-vanishing components of the Ricci tensor are $R_r^r = R_\varphi^\varphi = K$. Let us assume that in the four-dimensional spacetime these components of the Ricci tensor are also the only non-vanishing ones and use Einstein field equations $G_a^b = -8\pi \, T_a^b$. With $G_t^t = G_z^z = K$ and (25) we obtain:

$$F_z = \frac{1}{4} \left( \exp\left(-\nu\left(0,z\right)\right) - 1 \right). \tag{28}$$

Different "definitions" of the surface integral has been the origin of some confusion in the literature [7] [45]. The surface element used to derive (28) is $d\sigma \equiv \exp(\nu - 2\psi) \, r \, dr \, d\varphi$, which is the surface element of the 2-dimensional sub-manifold obtained by fixing $z$ and $t$ in the metric (1). On the other hand if one uses the element $\sqrt{|g|}drd\varphi = \exp(2\nu - 2\psi) \, r \, dr \, d\varphi$, as in Ref. [45], the compression force is $\frac{1}{4}\left(1 - \exp\left(\nu\left(0,z\right)\right)\right)$. Note that for small $\nu\left(0,z\right)$ the two forces coincide. Another source of confusion is the explicit form of the delta function for the distribution-value curvature. Of course the right use of distributions gives us the same expression (28), see for instance [46].

## 4 Two-body solutions and their forces

### 4.1 N point particles

The superposition of $N$ point particles (4)-(5) along the $z$-axis yield the following solution [5],

$$\psi = -\sum_{j=1}^{N} \frac{m_j}{R_j}, \tag{29}$$

$$\nu = -\frac{r^2}{2}\sum_j \frac{m_j^2}{R_j^4} + \sum_{j\neq k} \frac{m_j\, m_k}{(z_j - z_k)^2}\left[\frac{r^2 + (z-z_j)(z-z_k)}{R_j R_k} - 1\right], \quad (30)$$

$$R_j = \sqrt{r^2 + (z-z_j)^2}. \quad (31)$$

The integration constants were fixed to have at infinity an asymptotically flat space-time.

The case $N = 2$ promptly shows the strut singularity. The metric function $\nu$ has a non-zero value on the $z$-axis between the particles. According to the vacuum Einstein equation (3) and the elementary flatness boundary condition $\nu$ should vanish at $r = 0$. We get

$$\nu(0, z) = -4\frac{m_1 m_2}{(z_1 - z_2)^2} \quad (32)$$

for $z_1 < z < z_2$.

Let $D = z_2 - z_1$ be the coordinate distance between the particles. Then using (28) we get the compression force on the strut

$$F_z = \frac{1}{4}\left(\exp\left(4\frac{m_1 m_2}{D^2}\right) - 1\right) \approx \frac{m_1 m_2}{D^2} + 2\frac{(m_1 m_2)^2}{D^4}.$$

We say this is the module of the static attraction force between the particles. The Newtonian force is just the first term in the expansion above.

## 4.2 Two and N rods

Let us first consider de case of two massive rods (cf. Sec. 2.2), we find

$$\psi = \psi_1 + \psi_2, \quad (33)$$

$$\psi_j = -\frac{m_j}{L_j} \log \left| \frac{R_j^{(+)} + R_j^{(-)} - L_j}{R_j^{(+)} + R_j^{(-)} + L_j} \right|, \tag{34}$$

$$R_j^{(\pm)} = \sqrt{r^2 + \left(z_j^{(\pm)}\right)^2}, \tag{35}$$

$$z_j^{(\pm)} = z - \left(\zeta_j \mp \frac{L_j}{2}\right). \tag{36}$$

Then, the other metric function $\nu$ is given by

$$\nu[\psi] = \nu[\psi_1] + \nu[\psi_2] + 2\,\nu_{NL}[\psi_1, \psi_2],$$

$$\nu[\psi_j] = 2\left(\frac{m_j}{L_j}\right)^2 \log \left| \frac{\left(R_j^{(+)} + R_j^{(-)}\right)^2 - L_j^2}{4\,R_j^{(+)}\,R_j^{(-)}} \right|,$$

$$\nu_{NL}[\psi_1, \psi_2] = \frac{m_1 m_2}{L_1 L_2} \log \left| \frac{E\left(1^{(+)}, 2^{(-)}\right) E\left(1^{(-)}, 2^{(+)}\right)}{E\left(1^{(+)}, 2^{(-)}\right) E\left(1^{(-)}, 2^{(-)}\right)} \right|,$$

where we have introduced the notation

$$E\left(i^{(\pm)},\, j^{(\pm)}\right) = R_i^{(\pm)} R_j^{(\pm)} + z_i^{(\pm)} z_j^{(\pm)} + r^2,$$

for $i, j = 1, 2$.

The generalization for $N$ rods is straightforward. Note that, for $\zeta_i + \frac{L_i}{2} \leq z \leq \zeta_j - \frac{L_j}{2}$, $r = 0$, that is, the region at axis between the massive rods, the metric function $\nu(0, z)$ is not zero,

$$\nu(r = 0, z_{\text{between}}) \to 4\frac{m_1 m_2}{L_1 L_2} \log \left| \frac{(\zeta_1 - \zeta_2)^2 - (L_1 + L_2)^2}{(\zeta_1 - \zeta_2)^2 - (L_1 - L_2)^2} \right|.$$

Let $D$ be the (coordinate) distance between the center of the rods, that is, $|\zeta_1 - \zeta_2| = D + (L_1 + L_2)/2$, then the expression above simplifies to

$$\nu(r = 0, z_{\text{between}}) = 4\frac{m_1 m_2}{L_1 L_2} \log \left| 1 - \frac{L_1 L_2}{(D + L_1)(D + L_2)} \right|.$$

From (28) we get,

$$F_z = \frac{1}{4}\left(\left[\frac{(D+L_1)(D+L_2)}{D^2+D(L_1+L_2)}\right]^{4m_1m_2/(L_1L_2)} - 1\right),$$

$$\approx \frac{m_1m_2}{D^2} - \frac{m_1m_2}{D^3}\left\{L_1 + L_2 - \frac{\left(L_1^2 + \frac{3}{2}L_1L_2 + L_2^2 + 2\,m_1m_2\right)}{D}\right\},$$

where $L_1$ and $L_2$ are the rod length. This is the attraction force between two distorted black holes. It has been used successfully in the computation of gravitational radiation in the head-on collision of two black holes [27].

### 4.3 Two disks

The relations of the previous case can be used to calculate the force between two Morgan disks. The potential for two Morgan disks is obtained as the imaginary part of the superposition of two complex bars. We get

$$F_z = \frac{1}{4}\left(\left[\frac{D^2+(a_1+a_2)^2}{D^2+(a_1-a_2)^2}\right]^{m_1m_2/(a_1a_2)} - 1\right),$$

$$\approx \frac{mM}{D^2} - \frac{mM}{D^4}\left(a_1^2 + a_2^2 - 2m_1m_2\right).$$

### 4.4 True ring + particle

Now we shall consider the superposition

$$\psi = (\psi_{\text{Appell}} + \psi_{\text{Morgan}}) + \psi_{\text{CC}}, \tag{37}$$

where $\psi_{\text{CC}} = m/\sqrt{r^2 + (z-z_0)^2}$ is the potential of a Chazy-Curzon particle (4) with mass $m$ located at $r = 0$, $z = z_0$, and the expression in parenthesis is the Newtonian potential of a ring of radius $a$ with mass $M$ centered at $r = 0$ in the plane $z = 0$ obtained by the superposition of an Appell and an Morgan disks of mass $M/2$ each, as discussed above. The other metric function $\nu[\psi]$ can be easily computed using the identities of [21]. We find

$$\begin{aligned}\nu[\psi] &= \nu_{\text{Ring}} - \frac{m^2 r^2}{2 R_0} \\ &+ mM \left\{ \text{Re} \left[ \frac{1}{(z_0 - ia)^2} \left( \frac{r^2 + (z - ia)(z - z_0)}{R_a R_0} - 1 \right) \right] \right. \\ &\left. + \text{Im} \left[ \frac{2\mu_0}{a R_0 (\mu_0 - \mu_a)} \right] - \frac{2}{z_0^2 + a^2} \right\},\end{aligned} \qquad (38)$$

where $\nu_{\text{Ring}}$ is given by (23) and

$$R_0 = \sqrt{r^2 + (z - z_0)^2} \quad ; \quad \mu_0 = z_0 - z + R_0. \qquad (39)$$

The integration constant has been chosen such that $\lim_{r \to 0} \nu[\psi] = 0$ for $z > z_0$ and $z < 0$. The function $\nu$ is not a continuous function, for $0 < z < z_0$ we have

$$\lim_{r \to 0} \nu[\psi] = \frac{-2mM}{z_0^2 + a^2} - \frac{2mM(z_0^2 - a^2)}{(z_0^2 + a^2)^2} = -4mM \frac{z_0^2}{(z_0^2 + a^2)^2}. \qquad (40)$$

Therefore, there is a strut between the position of the particle and the center of the ring. Using the expression above into (28) we get the compression force on the strut between a particle and a ring (of mass $M$)

$$F_z = \frac{1}{4} \left( \exp \left[ 4mM \frac{D^2}{(D^2 + a^2)^2} \right] - 1 \right) \approx \frac{mMD^2}{(D^2 + a^2)^2} + 2 \frac{(mM)^2 D^4}{(D^2 + a^2)^4}.$$

where $D = z_0$ is the coordinate distance between the particle and the ring's center. In comparison, the Newtonian force between a particle and a ring, $-mMD/\left(D^2 + a^2\right)^{3/2}$, is less intense than the "general relativistic static" force above.

On the disk inside the ring the metric function $\nu$ is not continuous. That is $\nu(r, z = 0^+) \neq \nu(r, z = 0^-)$. The explicit single-valued expression of this jump is quite involved but the terms of (38) that contribute to it are just the ones proportional to $mM$. This jump in the metric function violates the Ricci flat structure of the spacetime – one concludes that it is no longer vacuum [47][48]. There is a membrane in this case.

We emphasize that these structures are consequences of the nonlinear character of the Einstein equations. Note that we do not have any singularity of the function $\psi$ at the inner part of the ring. This fact is shown in Fig. 2a where we present the potential of the superposition with $M = 3m = 1$. Nevertheless, for the function $\nu$, the singularities of the ring and the particle spread out in the form of a strut and a membrane by the nonlinear coupling between the sources singularities. This fact is depicted in Fig. 2b. Note the singularity on the disk limited by the ring. The strut singularity is not showed in Fig. 2b .

It is also instructive to see the functions $\nu$ associated with the Chazy-Curzon particle (Fig. 3a) and with the ring (Fig. 3b) separately. We observe singularities at the positions of the particle and ring only.

## 4.5 Struts, membranes and directional singularities

In all the superposition considered above, the bodies are associated to either singularities of the metric function $\psi$ or discontinuities of its first derivatives. Note that the singularities on $\psi$ also appear on $\nu$. To classify some of the singularities of the spacetime we present the projections of the Riemann curvature tensor along the natural orthogonal tetrad[4]. Using the vacuum field equations (2-3) the independent ones are given by

$$R_{\widehat{trtr}} = \kappa \left[ \psi_{rr} + \psi_r^2 \left( 2 - r\psi_r \right) - \psi_z^2 \left( 1 - 3r\psi_r \right) \right] \tag{41}$$

$$R_{\widehat{tztz}} = \kappa \left[ \psi_{zz} - \psi_r^2 \left( 1 - r\psi_r \right) + \psi_z^2 \left( 2 - 3r\psi_r \right) \right] \tag{42}$$

$$R_{\widehat{trtz}} = \kappa \left[ \psi_{zr} + 3\psi_r \psi_z \left( 1 - 3r\psi_r \right) + r\psi_z^3 \right] \tag{43}$$

where $\kappa \equiv \exp 2(\psi - \nu)$. For the solutions presented above, $(\psi - \nu) \to \pm \infty$, so $\kappa$ either diverges or vanishes where $\psi \to -\infty$ depending on the direction of approach. Thus by continuity of (41-43) the singularities of the Newtonian potential are physical directional singularities of the spacetime [36], but they can not be seen by an observer at infinity because the $g_{tt}$ component of the metric vanishes there [49][24][25].

Let us study the directional behavior of the ring singularity through the behavior of the scalar polynomial invariants of the curvature tensor. Us-ing the vacuum field equations (2-3) the simplest non-vanishing ones are $w_1 \equiv \frac{1}{8}C_{abcd}C^{abcd}$ and $w_2 \equiv -\frac{1}{16}C^{cd}_{ab}C^{ef}_{cd}C^{ab}_{ef}$ where $C^{abcd}$ is the Weyl trace-free tensor, which for vacuum solution coincides with the Riemann curvature tensor. After some algebraic manipulations [50] they are

$$\begin{aligned}w_1 &= 2\kappa^2\left[3\sigma\left(\psi_z^2 + r^2\psi_r^2\psi_z^2 - \psi_r/r - r\psi_r\sigma\right) + \psi_r^2\left(1 + 2r\nu_r + 3r\psi_r\right)\right.\\ &\quad + r^2\left(\psi_z^6 + \psi_r^6\right) + \psi_{rz}\left(6\psi_r\psi_z + \psi_{rz} - 2\left(\psi_z\nu_r + \psi_r\nu_z\right)\right)\\ &\quad \left. + \psi_{rr}\left(3\nu_r\left(1 - 2r\psi_r\right)/r - \psi_{zz} + 2\psi_z\nu_z\right)\right]\end{aligned}$$

$$\begin{aligned}w_2 &= \frac{3\kappa^3\left(\psi_r/r - \sigma\right)}{r^2}\left[\sigma\left(3r^3\psi_r^2\psi_z^2 + r\left(1 - 3r\psi_r\right)\sigma + 2\psi_r\right)\right.\\ &\quad - \psi_z^2\left(\nu_r + 3\psi_r\left(1 - 2r\psi_r\right)\right) + r^3\left(\psi_z^6 + \psi_r^6\right)\\ &\quad + r\psi_{rz}\left(2\psi_z\left(3\psi_r\left(1 - r\psi_r\right) + r\psi_z^2\right) + \psi_{rz}\right)\\ &\quad \left. + \psi_{rr}\left(3\nu_r\left(1 - 2r\psi_r\right) + 4r^2\psi_r^3 - r\psi_{zz}\right)\right]\end{aligned}$$

where $\sigma \equiv \psi_z^2 + \psi_r^2$. In the neighborhood or the ring the metric functions are given by

$$\psi \approx -M\sqrt{\frac{\sqrt{(r-a)^2 + z^2} + r - a}{4a(r-a)^2}}$$

$$\nu \approx \frac{M^2}{2a^2}\log\left[\frac{2(r-a)^2 + z^2}{a}\frac{}{|r-a|}\right]$$

---
[4]Distribution like singularities can not be studied in this manner.

Let the approach to the ring be through the straight paths $r = a + \varepsilon$ and $z = \tan\alpha\,\varepsilon$ for the parameter $\varepsilon$ and an angle $\alpha$ such that $|\alpha| < \pi/2$. The leading term for $w_1$ and $w_2$ as $\varepsilon \to 0$ are:

$$\lim_{\varepsilon \to 0} w_1 \approx \frac{m^6 \, 4^{-\frac{m^2}{4}}}{16384 \, a^4} \frac{(\cos \alpha)^{m^2-3} (2 - \cos^2 \alpha)^3}{|\varepsilon|^{m^2/2} \, \varepsilon^9 \exp\left(m\sqrt{\left(\frac{|\varepsilon|}{\cos \alpha} + \varepsilon\right) \varepsilon^{-2}}\right)}$$

$$\lim_{\varepsilon \to 0} w_2 \approx -\frac{3m^8 \, 4^{-3m^2/8}}{1048576 \, a^7} \frac{(\cos \alpha)^{3m^2/2-4} (2 - \cos^2 \alpha)^4}{|\varepsilon|^{3m^2/4} \, \varepsilon^{12} \exp\left(\frac{3m}{2}\sqrt{\left(\frac{|\varepsilon|}{\cos \alpha} + \varepsilon\right) \varepsilon^{-2}}\right)}$$

where $m \equiv M/a$. We find the directional behavior of $w_1$ and $w_2$ as

$$\lim_{\varepsilon \to \pm 0} \left\{ \begin{array}{c} w_1 \\ w_2 \end{array} \right. = \left\{ \begin{array}{lll} 0 & \text{for} \quad \alpha = 0 & \text{and} \quad \varepsilon \to +0 \\ -\infty & \text{for} \quad \alpha = 0 & \text{and} \quad \varepsilon \to -0 \\ 0 & \text{for} \quad \alpha \to \pm\pi/2 & \text{and} \quad \varepsilon \to \pm 0 \end{array} \right.$$

Thus, the ring is singular when one approaches it from its interior. Otherwise the spacetime is extendible [35].

The discontinuity of the metric function $\nu$ gives rise to the strut and the membrane as we consider superposition of bodies. The former is a conic-like singularity and the later is a combination of delta function and its derivatives with support on the inner part of the ring. Thus, for the inner part of the ring we have a rather strong singularity with a multi-polar character. One can show that the strut is related to a non-vacuum region with a delta function energy-momentum tensor [48][44]. Another example of the appearance of a membrane in the inner part of the ring has been given recently [51].

The conic structure studied here are gravitationally inert. In general, one has that the associated Newtonian density of this objects is null. The associated gravitational Newtonian density is defined as,

$$\rho_N = \rho + p_1 + p_2 + p_3,$$

where $\rho$ and the $p_a$ are the eigenvalues of the EMT [52]. For the struts computed in this section we have $-\rho = p_z > 0$, the same equation of state of cosmic strings [53].

# 5 Conclusion

Thus the Weyl spacetime solutions have very peculiar features. We emphasized the superpositions solutions in which the singularities in one of the metric functions, $\psi$ is somehow spread over other regions by the non-linearity of the other metric function $\nu$.

In the superposition of two contiguous bodies a strut singularity appears between them. For the superposition of a ring and another body, besides the strut there appears a membrane-like singularity. Therefore our conclusions of the letter [20] still hold in the sense that there appears a strut and a membrane in order to hold a point and a ring in equilibrium. The membrane is not an usual matter even in the sense of distribution valued tensors, but it seems that this is the price of keeping the ring in equilibrium with a point particle. We believe that the appearance of membranes to equilibrate hollow bodies is a general feature of the Weyl solutions. We have studied a variety of situations wherein the same phenomena occur.

The forces we obtained in this paper are just the compression forces on the strut needed to achieve the static configuration of the solution with two bodies. But, in a sense of action-a-distance, it does give us the intensity of, say, the static forces between the two axisymmetric bodies.

We want to thank professor W.B. Bonnor (Q.M.W. College) for his interest in this research and an anonymous referee for pointing out some misleading equations and relevant references. We also want to thank CNPq and FAPESP for financial support.

Figure 1: a) The Appell potential is shown for $a = 2M = 1$. The equipotentials for $r \leq 1$ and $z$ near the plane $z = 0$ tell us that there is a structure inside the Appell ring. b) For the same values of the parameters we show the equipotentials for a Morgan disk. c) The superposition of the previous two potentials shows that we have a ring with no structure inside the ring.

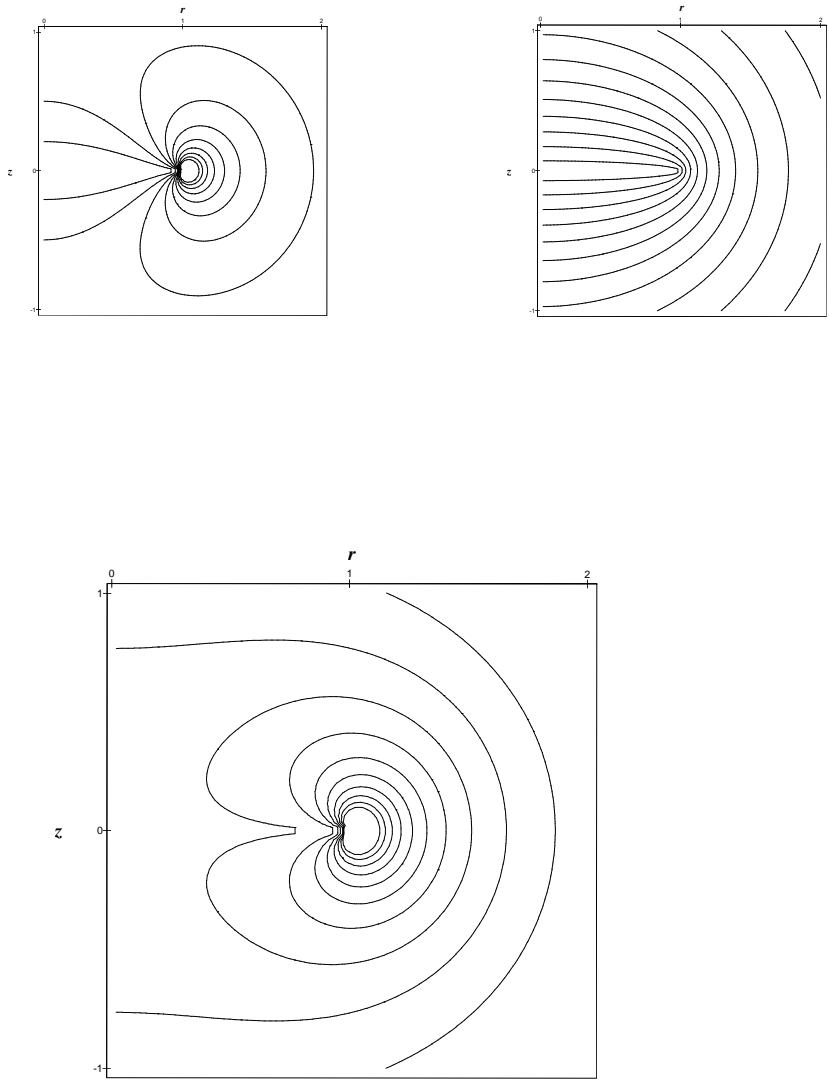

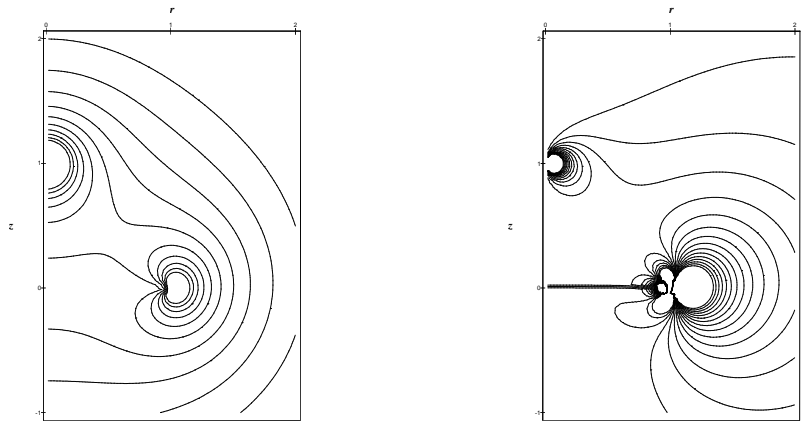

Figure 2: a) The potential of the superposition of a Chazy-Curzon particle of mass $m = 1/3$ located on the $z$-axis at $z = 1$ with a ring of $M = a = 1$ located on the plane $z = 0$ is shown. b) We have that the $\nu$ function of the superposition is singular on the ring and the particle, and it is also discontinuous on the disk whose boundary is the ring.

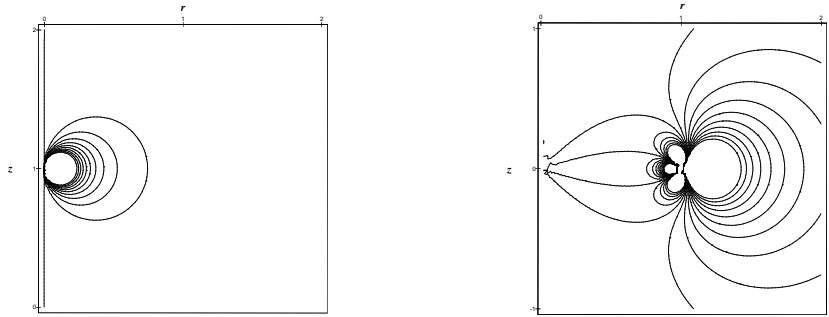

Figure 3: In a) and b) we show the functions $\nu$ for the Chazy-Curzon particle and the ring, respectively. Note that these functions are singular only on the position of the particle and the ring and they are continuous everywhere else. In particular the plane enclosed by the ring is regular.